# EXACT SOLUTION OF TWO-DIMENSIONAL SCREENED DONOR STATE IN A MAGNETIC FIELD

Le Van Hoang, Le Tran The Duy
Hoang Do Ngoc Tram, Ngo Dinh Nguyen Thach, Le Thi Ngoc Anh

Department of Physics, HCMC University of Pedagogy
280, An Duong Vuong, Dist. 5, HCM city

**Abstract.** The use of Levi-Civita transformation allows us to formulate the problem of two-dimensional screened donor states in a magnetic field as that of two-dimensional anharmonic oscillator. Therefore, the operator method can be directly used for the first problem and the exact solutions of Schrödinger equation are obtained correspondingly. In our approach, wave-functions are constructed in the representation of annihilation and creation operators, which permits one to use purely algebraic method in further calculations of other characteristics. The considered problem is related to the motion of 2D electron gas in GaAs/AlGaAs multiple-quantum well structures with the presence of a magnetic field, which continues to provide new and fascinating phenomena.

## 1. INTRODUCTION

Electron gas in the two-dimensional crystal structure has been intensively investigated in the last few years [1-9]. Saying, the multi-layer semiconductor structure in which GaAs regions act as quantum wells for the conduction electron while $Al_xGa_{1-x}As$ ($x \leq 0.45$) regions play the role of quantum barriers, is the majority super-lattice considered presently [1]. A lot of new and fascinating phenomena relating to the motion of two-dimensional electron gas in GaAs/AlGaAs multiple-quantum well structures with the presence of magnetic field [2-4] continue to be discovered. So, one of the most interesting problems continuatively considered till now is the donor states in a magnetic field [5-9].

In atomic unit system, the Schrödinger equation of screened donor states in a magnetic field can be written as follows:

$$\hat{H}\,\Psi(\mathbf{r}) = E\,\Psi(\mathbf{r}), \qquad (1)$$

$$\hat{H} = -\frac{1}{2}\left(\frac{\partial^2}{\partial x^2} + \frac{\partial^2}{\partial y^2}\right) - \frac{1}{2}i\gamma\left(x\frac{\partial}{\partial y} - y\frac{\partial}{\partial x}\right) + \frac{1}{8}\gamma^2\left(x^2 + y^2\right) - \frac{e^{-\lambda r}}{r}. \qquad (2)$$

The energy unit is the effective Rydberg constant $R^* = m*e^4/2\hbar^2\varepsilon^2$; the coordinates are measured in unit of the effective radius Bohr $a^* = \varepsilon\hbar^2/e^2 m^*$ and the dimensionless parameter $\gamma$ is defined by the formula $\gamma = \hbar\omega_c/2R^*$ with the cyclotron frequency $\omega_c = eB/m^*c$ and the magnetic intensity $B$. Here, $m*$, $\varepsilon$ are the electron effective mass and the static dielectric constant respectively.

Equation (1)-(2) has two points different from almost works in the similar topic. (i) The interaction between quantum hole and donor electron is via Screened Coulomb potential of Yukawa type with the positive parameter λ. This parameter depends on

Email: hoanglv@hcmup.edu.vn

many factors of the system and should be found by comparing experimental data with theoretical computation results. (ii) In the case of using semiconductor structure GaAs/$Al_xGa_{1-x}$As and a laboratorial magnetic field, the Coulomb energy has the same scale with the magnetic interaction. In other words, besides the case of a weak magnetic field, solutions in the diapason $\gamma \approx 1$ are interesting in practice too. Therefore, in equation (1)-(2) the magnetic intensity $\gamma$ is considered arbitrary.

Evidently, the usual methods of the perturbation theory can not be directly used in these cases. So, in this paper, we suggest using the operator method (OM) to solve the equation (1)-(2). This method was first built and given in the work [10] as a non-perturbation method for solving the Schrödinger equation of quantum systems with external interaction of arbitrary intensity. Till now, OM remains useful for the majority problems of atomic calculations; solid state physics and quantum field theory (see [11-12]).

## 2. RELATION WITH TWO-DIMENSIONAL HARMONIC OSCILLATOR

The equation (1)-(2) will become simpler after the substitution of the Levi-Civita transformation [13-14] as follows:

$$\begin{cases} x = u^2 - v^2 \\ y = 2uv. \end{cases} \quad (3)$$

with properties: $dx\,dy = 4(u^2 + v^2)\,du\,dv$, $r = \sqrt{x^2 + y^2} = u^2 + v^2$. Since Jacobian of transformation (3) is not a constant, the weight $4(u^2 + v^2)$ will appear in the expression of scalar product of two state-vectors in the xy- space when transforming into uv- space. Consequently, if any $\hat{K}$ is a Hermitian operator in xy- space then the operator $\tilde{K} = 4(u^2 + v^2)\hat{K}$ will be Hermitian in uv- space accordingly. So, for conserving the Hermitian property of Hamiltonian after the transformation (3), the equation (1) should be rewritten in the form:

$$r(\hat{H} - E)\Psi(\mathbf{r}) = 0.$$

In the uv- space, this equation becomes

$$\tilde{H}\,\Psi(u, v) = 0, \quad (4)$$

with the Hermitian Hamiltonian

$$\tilde{H} = -\frac{1}{8}\left(\frac{\partial^2}{\partial u^2} + \frac{\partial^2}{\partial v^2}\right) - (E + \frac{\gamma}{2}\hat{L}_z)(u^2 + v^2) + \frac{\gamma^2}{8}(u^2 + v^2)^3 - \exp\{-\lambda(u^2 + v^2)\}. \quad (5)$$

It is easy to see that (5) is the Hamiltonian of anharmonic oscillator in two-dimensional space. It means that we have transformed the complicated problem of the electron motion in the electromagnetic field into the simpler, well-considered problem in Quantum Mechanics [15].

Energy E is no longer an eigen-value of equation (4). In fact, E just plays a role of parameter and for equation (4) eigen-values are always equal to zero. For convenient, let us consider the eigen-value Z of equation (4) as a function of parameter E and from the equation $Z(E) = 0$, we will obtain the quantity of energy E. Besides, the angular momentum operator $\hat{L}_z$, having the form:

$$\hat{L}_z = \frac{i}{2}\left(u\frac{\partial}{\partial v} - v\frac{\partial}{\partial u}\right)$$

in uv- space, commutes with the Hamiltonian (5). It means the angular momentum corresponding to the z direction is an integral of motion in the screened Coulomb potential plus a magnetic field. This fact should be later taken into consideration while solving the equation (4).

## 3. OPERATOR METHOD OF SOLVING SCHRÖDINGER EQUATION

To solve the equation (4)-(5) by the operator method we follow four steps.

**Step 1:** Firstly, we transform equation (4)-(5) into the representation of annihilation and creation operators by using the following definitions:

$$\hat{a}(\omega) = \sqrt{\frac{\omega}{2}}\left(\xi + \frac{1}{\omega}\frac{\partial}{\partial\xi^*}\right), \quad \hat{a}^+(\omega) = \sqrt{\frac{\omega}{2}}\left(\xi^* - \frac{1}{\omega}\frac{\partial}{\partial\xi}\right),$$

$$\hat{b}(\omega) = \sqrt{\frac{\omega}{2}}\left(\xi^* + \frac{1}{\omega}\frac{\partial}{\partial\xi}\right), \quad \hat{b}^+(\omega) = \sqrt{\frac{\omega}{2}}\left(\xi - \frac{1}{\omega}\frac{\partial}{\partial\xi^*}\right),$$

(6)

where the complex coordinates are defined by equations: $\xi = u + iv$, $\xi^* = u - iv$. Operators (6) satisfy the standard commutation correlations:

$$\hat{a}\,\hat{a}^+ - \hat{a}^+\,\hat{a} = 1, \qquad \hat{b}\,\hat{b}^+ - \hat{b}^+\,\hat{b} = 1 \qquad (7)$$

(The other commutators are equal to zero). Here, the free parameter ω is a real positive number defined in the next steps. After putting (6), equation (4)-(5) has the form:

$$\left\{\left(\frac{\omega}{4} - \frac{2E + m\gamma}{4\omega}\right)(\hat{a}^+\hat{a} + \hat{b}^+\hat{b} + 1) - \left(\frac{\omega}{4} + \frac{2E + m\gamma}{4\omega}\right)(\hat{a}^+\hat{b}^+ + \hat{a}\,\hat{b})\right.$$

$$+ \frac{\gamma^2}{64\omega^3}(\hat{a}^+\hat{b}^+ + \hat{a}\,\hat{b} + \hat{a}^+\hat{a} + \hat{b}^+\hat{b} + 1)^3$$

$$\left.- \exp\left\{-\frac{\lambda}{2\varpi}(\hat{a}^+\hat{b}^+ + \hat{a}\,\hat{b} + \hat{a}^+\hat{a} + \hat{b}^+\hat{b} + 1)\right\}\right\}|\Psi\rangle = Z|\Psi\rangle. \quad (8)$$

In the equation (8), the exponential operator can be reduced in the normal form as

$$\hat{A} = \exp\left\{-\frac{\lambda}{\lambda + 2\omega}\hat{a}^+\hat{b}^+\right\}\exp\left\{\ln\left(\frac{2\omega}{\lambda + 2\omega}\right)(\hat{a}^+\hat{a} + \hat{b}^+\hat{b} + 1)\right\}\exp\left\{-\frac{\lambda}{\lambda + 2\omega}\hat{a}\,\hat{b}\right\}$$

in the meaning that the annihilation operators locate in the right-hand side and all the creation ones locate in the left-hand side. Therefore, it is suitable for algebraic calculations.

**Step 2:** We now separate the Hamiltonian of equation (8) into two parts as follows:

$$\widetilde{H} = \widetilde{H}_0 + \beta \widetilde{V} , \tag{9}$$

where $\widetilde{H}_0$ contains only the terms commuting with the neutral operators $\hat{a}^+\hat{a}$ , $\hat{b}^+\hat{b}$ :

$$\widetilde{H}_0 = \left( \frac{\omega}{4} - \frac{2E + m\gamma}{4\omega} \right)(\hat{a}^+\hat{a} + \hat{b}^+\hat{b} + 1) +$$

$$+ \frac{\gamma^2}{64\omega^3}(\hat{a}^+\hat{a} + \hat{b}^+\hat{b} + 1)[(\hat{a}^+\hat{a} + \hat{b}^+\hat{b} + 1)(\hat{a}^+\hat{a} + \hat{b}^+\hat{b} + 4) + 6\hat{a}^+\hat{a}\hat{b}^+\hat{b} + 2]$$

$$- \exp\left\{ \ln\left( \frac{2\omega}{\lambda + 2\omega} \right)(\hat{a}^+\hat{a} + \hat{b}^+\hat{b} + 1) \right\} \sum_{k=0}^{+\infty} \frac{1}{(k!)^2} \left( \frac{-\lambda}{2\omega} \right)^{2k} (\hat{a}^+\hat{b}^+)^k (\hat{a}\hat{b})^k . \tag{10}$$

The remaining part $\widetilde{V} = \widetilde{H} - \widetilde{H}_0$ is considered as perturbative term. The parameter β is given to indicate that the term $\widetilde{V}$ in (9) is "smaller" than $\widetilde{H}_0$. The parameter $\omega$ will be chosen in the way that the perturbative condition:

$$\left\| \widetilde{H}_0 \right\| >> \left\| \widetilde{V} \right\| \tag{11}$$

can be satisfied.

**Step 3:** We will solve the equation:

$$\widetilde{H}_0(\hat{a}^+\hat{a}, \hat{b}^+\hat{b}) \left| \Psi^{(0)} \right\rangle = Z^{(0)} \left| \Psi^{(0)} \right\rangle , \tag{12}$$

for the solutions in zero order of approximation. It is easy to see that they are the eigen-vectors of the operators $\hat{a}^+\hat{a}$ , $\hat{b}^+\hat{b}$ . In other words, solutions of (12) are the wave-vectors of two-dimensional harmonic oscillator. Moreover, the conservation of the angular momentum in z direction leads to an additional equation:

$$(\hat{a}^+\hat{a} - \hat{b}^+\hat{b}) \left| \Psi^{(0)} \right\rangle = 2m \left| \Psi^{(0)} \right\rangle , \tag{13}$$

where the angular momentum operator in the representation of the operators (6) $\hat{L}_z = \frac{1}{2}(\hat{a}^+\hat{a} - \hat{b}^+\hat{b})$ has been used. Finally, the solutions of the equations (12) and (13) can be obtained in the form:

$$\left| n(m) \right\rangle = \frac{1}{\sqrt{(n-m)!(n+m)!}} (\hat{a}^+)^{n-m}(\hat{b}^+)^{n+m} \left| 0(\omega) \right\rangle \tag{14}$$

with n is the quantum principal number ( n = 0,1,2,…. ) and m is the azimuthal number satisfying $-n \le m \le n$ . Here, the vacuum-state is defined by equations:

$$\hat{a}(\omega) \left| 0(\omega) \right\rangle = 0, \quad \hat{b}(\omega) \left| 0(\omega) \right\rangle = 0. \tag{15}$$

For the use in latter, we write some important formulae:

$$\hat{a}^+\hat{b}^+ \left| n(m) \right\rangle = \sqrt{(n+1)^2 - m^2} \left| n+1 \ (m) \right\rangle ,$$

$$\hat{a}\hat{b} \left| n(m) \right\rangle = \sqrt{n^2 - m^2} \left| n-1 \ (m) \right\rangle ,$$

$$(\hat{a}^+\hat{a} + \hat{b}^+\hat{b})\ |n(m)\rangle = 2n\ |n(m)\rangle .$$

By putting the state-vector (14) into (12) and using the equation $Z^{(0)}(E_{nm}^{(0)}) = 0$ we obtain the expression of energy in the given state respectively:

$$E_{nm}^{(0)} = \frac{\omega^2 - \mathrm{m}\gamma}{2} + \frac{\gamma^2}{16\omega^2}(5\mathrm{n}^2 + 5\mathrm{n} + 3 - 3\mathrm{m}^2) - \frac{2\omega}{(2\mathrm{n}+1)(\alpha+1)^{2n+1}}F_0\big(n,m,\alpha^2\big), \quad (16)$$

where $\alpha = \dfrac{\lambda}{2\omega}$ ; $F_0(n,m,x)$ is the confluent hyper-geometric function with definition:

$$F_j(n,m,x) = {_2F_1}(m-n,-m-n;j+1;x) = \sum_{k=0}^{n-|m|}\frac{j!(n-m)!(n+m)!}{k!(k+j)!(n-m-k)!(n+m-k)!}x^k .$$

Parameter $\omega$ can be defined by the condition:

$$\frac{\partial E_{nm}^{(0)}}{\partial \omega} = 0 . \qquad (17)$$

The problem of defining parameter $\omega$ has been considered in some works (see for example [12]) and it is proved that the condition (17) conforms to the condition (11) and thus provides for obtaining the good results just in the zero order of approximation. In our case, condition (17) leads to the equation:

$$\omega^4 - \frac{\gamma^2}{8}(5\mathrm{n}^2 + 5\mathrm{n} + 3 - 3\mathrm{m}^2)$$
$$- 2\omega^3\frac{(1+2(n+1)\alpha)F_0\big(n,m;\alpha^2\big) - 2(n^2-m^2)\,(1+\alpha)\alpha^2 F_1\big(n-1,m,\alpha^2\big)}{(2n+1)\big(1+\alpha\big)^{2n+2}} = 0 , \quad (18)$$

which has the positive real solution for value $\omega$. By putting this solution into (16) we obtain the analytical expression for the energy in zero approximation. It should be noted that for the equations (16), (18) the symbolic computation can be used with the help of Mathematica. In the case of a super strong magnetic field, the asymptotic behavior of the wave-functions

$$e^{-\upsilon(x^2+y^2)}$$

when $\gamma \gg 1$ has to be considered additionally. Fig. 1 shows the magnetic intensity appendage of the energy obtained from solutions of equations (16), (18) for the ground state and some low excited states. For comparison, in Fig.1 the exact results obtained in the next section for the same states are shown also.

**Step 4:** We now construct the scheme for exact numerical solutions. Since the set of state-vectors (14) with the fixing value $m$ and varying value $n \geq |m|$ have all properties of the so-called complete system, the exact wave-vector can be expressed via the series power of these states as follows:

$$\big|\Psi_{n(m)}\big\rangle = \big|\mathrm{n(m)}\big\rangle + \sum_{\substack{k=|m| \\ k \neq n}}^{\infty} C_k\big|k(m)\big\rangle , \qquad (19)$$

with the real coefficients $C_k$ $(k = |m|, |m| + 1, \ldots; k \neq n)$. Putting (19) into equation (8), and then comparing the coefficients appeared in each state-vector of (14) we obtain the equations:

$$Z_n = \widetilde{H}_{nn} + \beta \sum_{\substack{k=|m| \\ k \neq n}}^{+\infty} C_k \, \widetilde{H}_{nk} \quad , \tag{20}$$

$$\left(Z_n - \widetilde{H}_{jj}\right) C_j = \beta \, \widetilde{H}_{jn} + \beta \sum_{\substack{k=|m| \\ k \neq n \\ k \neq j}}^{+\infty} C_k \, \widetilde{H}_{jk} \quad , (j = |m|, |m| + 1, \ldots \quad ; j \neq n). \tag{21}$$

Here, the matrix elements of the operator $\widetilde{H}$ according to state-vectors (14) are easy to calculate algebraicly with the use of commutative correlations (7) and the equations (15). When calculating, the most difficult term seems to be related to the matrix elements of the operator $\hat{A}$. However, the success in constructing the normal form of this operator as shown below allows us to find these elements as:

$$A_{n,n+j} = \left\langle n(m) \middle| \hat{A} \middle| n + j(m) \right\rangle = \frac{(-\alpha)^j}{(1+\alpha)^{2n+j+1}} \sqrt{\frac{(n+j-m)!(n+j+m)!}{(n-m)!(n+m)!}} \, F_j\left(n, m, \alpha^2\right) \,,$$

with $j \geq 0$. We write some non-zero matrix elements as follows:

$$\widetilde{H}_{nn} = \left(\frac{\omega}{4} - \frac{2E + m\gamma}{4\omega}\right)(2n+1) + \frac{\gamma^2}{32\omega^3}(2n+1)(5n^2 + 5n + 3 - 3m^2) - A_{nn} \,,$$

$$\widetilde{H}_{n,n+1} = \left(-\frac{\omega}{4} - \frac{2E + m\gamma}{4\omega} + \frac{3\gamma^2}{64\omega^3}(5n^2 + 10n + 6 - m^2)\right)\sqrt{(n+1)^2 - m^2} - A_{n,n+1} \,,$$

$$\widetilde{H}_{n,n+2} = \frac{3\gamma^2}{64\omega^3}(2n+3)\sqrt{(n+1)^2 - m^2}\sqrt{(n+2)^2 - m^2} - A_{n,n+2} \,,$$

$$\widetilde{H}_{n,n+3} = \frac{\gamma^2}{64\omega^3}\sqrt{(n+1)^2 - m^2}\sqrt{(n+2)^2 - m^2}\sqrt{(n+3)^2 - m^2} - A_{n,n+3} \,,$$

$$\widetilde{H}_{n,n+j} = -A_{n,n+j} \quad (j \geq 4). \tag{22}$$

Besides (22), the other non-zero matrix elements can be calculated from the symmetrical property $\widetilde{H}_{nk} = \widetilde{H}_{kn}$.

Now, let us solve the equation system (20)-(21) by the perturbation method using expansion of power series of the perturbation parameter $\beta$. As a result, for s-order of approximation, we have:

$$Z_n^{(s)} = \sum_{k=0}^{s} \Delta Z_n^{(k)} \, , \ C_j^{(s)} = \sum_{k=0}^{s} \Delta C_j^{(k)} \tag{23}$$

where $\quad \Delta Z_n^{(0)} = \widetilde{H}_{nn} \, , \ \Delta Z_n^{(1)} = 0 \, , \ \Delta C_j^{(0)} = 0 \, , \ \Delta C_j^{(1)} = \dfrac{\widetilde{H}_{jn}}{Z_n^{(0)} - \widetilde{H}_{jj}} \, ,$

$$\Delta Z_n^{(k)} = \sum_{\substack{j=|m| \\ j \neq n}}^{+\infty} \Delta C_j^{(k-1)} \widetilde{H}_{nj} \ (k \geq 2) \, , \tag{24}$$

$$\Delta C_i^{(k)} = \sum_{\substack{j=|m| \\ j \neq n \\ j \neq j}}^{+\infty} \frac{\widetilde{H}_{ij}}{\widetilde{H}_{nn} - \widetilde{H}_{ii}} \Delta C_j^{(k-1)} - \sum_{t=1}^{k-1} \frac{\Delta Z_n^{(k-t-1)}}{\widetilde{H}_{nn} - \widetilde{H}_{ii}} \Delta C_i^{(t)} \quad (k \geq 2). \tag{25}$$

By putting solutions of the recurrent equations (24)-(25) into (23), we define the coefficients $C_j$ and $Z_n$ in any given s – order of approximation. Accordingly, the energy $E_{nm}^{(s)}$ can be acquired in the same approximation order. The numerical results approve that the series

$$E_{nm}^{(0)}, E_{nm}^{(1)}, E_{nm}^{(2)}, \ldots, E_{nm}^{(s)}, \ldots$$

are rapidly convergent to a certain value $E_{nm}^{T}$, which may be considered as an exact numerical solution of the equation (1)-(2). Exact results for ground state and for some first excited states are given in tab. 1, tab. 2 and fig. 1, fig. 2, fig. 3 in the case of $\lambda = 0$.

**Tab 1:** The magnetic intensity dependence of energy levels - exact solutions in comparison with the zero order approximation ones

| $\gamma$ | 1s | | 2s | |
|---|---|---|---|---|
| | $E^{(o)} + \Delta E^{(2)}$ | $E^{T}$ | $E^{(o)} + \Delta E^{(2)}$ | $E^{T}$ |
| 0.1 | -1.999531227300 | -1.999531492082 | -0.189425392478 | -0.205286888607 |
| 0.5 | -1.988270414800 | -1.988425281714 | 0.040830382398 | 0.040317408544 |
| 1.0 | -1.953113823263 | -1.955159683247 | 0.457376245815 | 0.494679638837 |
| 2.0 | -1.816655524914 | -1.836207439051 | 1.428479429477 | 1.576895542024 |
| 5.0 | -1.042926239466 | -1.226356637452 | 4.718418831904 | 5.271097760845 |
| 10 | 0.829712394045 | 0.184843852730 | 10.628237966548 | 11.883586469278 |

**Tab. 2:** The magnetic intensity dependence of energy levels - exact solutions

| γ | 3s | 3p+ | 3p- | 3d+ | 3d- |
|------|---|-----|-----|-----|-----|
| 0.10 | 0.001634680897 | 0.043576080008 | -0.056423919992 | 0.071795722015 | -0.128204277985 |
| 0.50 | 0.676770387524 | 0.828653444114 | 0.328653444114 | 0.892438209302 | -0.107561790698 |
| 1.00 | 1.676970710757 | 1.918103969141 | 0.918103969141 | 2.005694128790 | 0.005694128790 |
| 2.00 | 3.828388290161 | 4.191127807757 | 2.191127807756 | 4.311633609260 | 0.311633609260 |
| 5.00 | 10.651147510593 | 11.240695516090 | 6.240695516090 | 11.425895187055 | 1.425895187055 |
| 10.00 | 22.39839272371 | 23.232965535868 | 13.232965535869 | 23.490841925399 | 3.490841925399 |

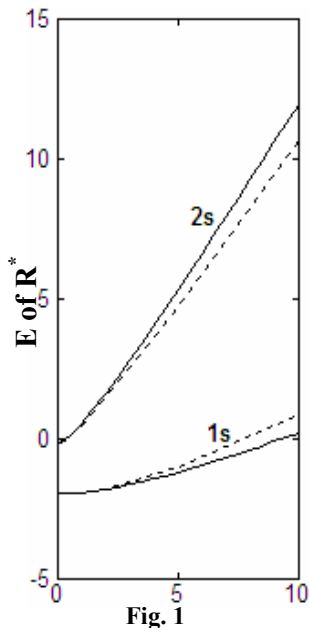
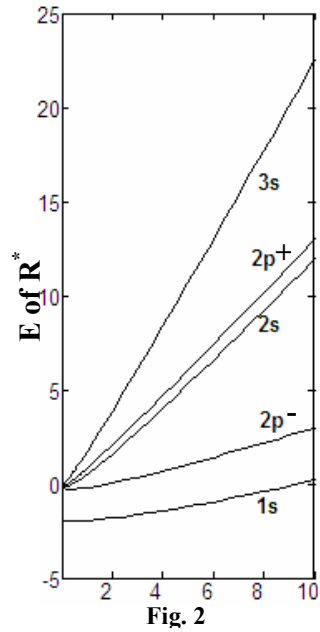
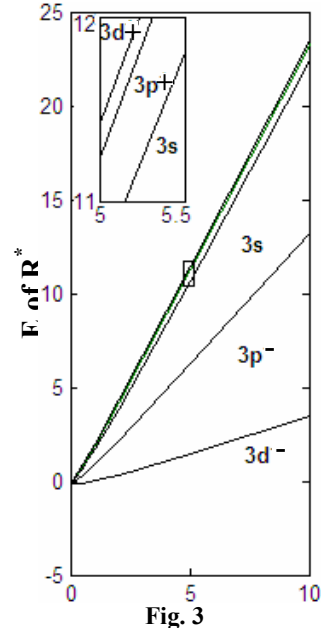

**Fig. 1:** The magnetic intensity dependence of energy levels - exact solutions in comparison with the zero order approximation ones (the dot-dashed lines).

**Fig. 2:** The magnetic intensity dependence of energy levels - exact solutions for 1s, 2s, 2p+, 2p- and 3s states.

**Fig. 3:** The magnetic intensity dependence of energy levels - exact solutions for 3s, 3p+, 3p-, 3d+ and 3d- states.

## 4. CONCLUSION

Consequently, we have described the effective method for solving the problem of screened donor states in a magnetic field with the arbitrary intensity. The method is simple since we can use purely algebraic calculation. The obtained analytical equations are suitable for using symbolic computation such as Mathematica. A part of the results have been reported in the publication [9], the more detailed and complete data obtained by this method will be analyzed and reported in our other work. The method can be recommended for other problems, for example, the problem of exciton in carbon nanotubes (see [17]).

## 5. ACKNOWLEDGEMENT

This work is supported by grant CS.2004.23.59 of HCMC University of Pedagogy.